\documentstyle[prl,aps,floats,psfig]{revtex}

\begin{document}

\widetext
\draft
\twocolumn[\hsize\textwidth\columnwidth\hsize\csname @twocolumnfalse\endcsname
\title{Size Effects in Carbon Nanotubes} 

\author{
C.-H.\ Kiang$^1$, 
M.\ Endo$^2$, P.\ M.\ Ajayan$^3$
G.\ Dresselhaus$^4$, and M.~S.\ Dresselhaus$^5$
}
\address{
$^1$Department of Chemistry and Biochemistry,
University of California, Los Angeles, CA 90095\\
$^2$Department of Electrical and Electronics Engineering,
Faculty of Engineering,
Shinshu University, Nagano 380, Japan\\
$^3$Department of Materials Science and Engineering,
Rensselaer Polytechnic Institute, Troy, NY 12180\\
$^4$Francis Bitter Magnet Laboratory,
Massachusetts Institute of Technology, Cambridge, MA 02139\\
$^5$Department of Electrical Engineering and
Computer Science and Department of Physics \\
Massachusetts Institute of Technology, Cambridge, MA 02139
}

\maketitle

\begin{abstract}
The inter-shell spacing of multi-walled carbon nanotubes was
determined by analyzing the high resolution 
transmission electron microscopy images of these
nanotubes.  For the nanotubes that were studied, 
the inter-shell spacing ${\hat{d}_{002}}$ 
is found to range from 0.34 to 0.39 nm, increasing   
with decreasing tube diameter.
A model based on the results from real space image analysis
is used to explain the variation in inter-shell spacings
obtained from reciprocal space periodicity analysis. 
The increase in inter-shell spacing with decreased nanotube diameter
is attributed to the high curvature, resulting in an increased   
repulsive force, associated with the decreased diameter 
of the nanotube shells.
\end{abstract}

]

The discovery of multi-walled carbon nanotubes \cite{Iijima91a}
and single-walled carbon nanotubes \cite{Iijima93b,Kiang93b}
has prompted numerous studies of the structure, properties
\cite{Dresselhaus95a,Mint95a,Ebbesen92a,Endo92a,Zettl95a},
and potential applications 
\cite{Iijima93a,Smalley95a,Ruoff95a}
of these exotic materials.
For example, nanotubes are expected to have a high strength-to-weight 
ratio \cite{Ruoff95a}, which is advantageous in advanced composites 
to be used in high performance materials such as aircraft frames.
The small dimension of the tubes shows promise for use as a  
gas absorption medium \cite{Gubbins95a,Kiang97a},
a field emitter for use in flat-panel displays \cite{Ugarte95a},  
and a nanoscale electronic 
device \cite{Cohen96a,Srivastava97a,Lieber98a,Dekker98a}.
Most of the anticipated properties and applications are 
based on theoretical calculations for idealized tube structures
\cite{Oshiyama92a,Broughton92a,Akagi95a}.
A precise knowledge of the structure of real nanotubes and 
the interactions between them
is essential for a reliable prediction of the potential applications.

Structural studies of carbon nanotubes have relied heavily
on X-ray diffraction \cite{SaitoY93d}, scanning tunneling microscopy
\cite{Sattler93a}, and, predominately, 
high resolution transmission electron microscopy (HRTEM) and
electron diffraction \cite{Iijima91a,Amel93a,Cowley,Kiang94a}.
Carbon nanotubes are composed of concentric cylindrical graphene
tubules, each with a structure similar to that of a 
rolled-up graphene sheet \cite{Dresselhaus95a}.
Iijima first showed with transmission electron microscopy that the 
inter-shell spacing of carbon nanotubes is about 0.34 nm 
\cite{Iijima91a}, which
was later confirmed by Zhang {\em et.\ al}\ with an electron diffraction
study \cite{Amel93a}.
Saito {\em et.\ al}\ used powder X-ray diffraction to 
determine the lattice parameters of a bulk nanotube
sample \cite{SaitoY93d}, and they concluded that the average
inter-shell spacing is 0.344 nm.  
Bretz {\em et.\ al}, on the other hand, 
obtained a 0.375 nm spacing \cite{Bretz94a},
and Sun {\em et al.} obtained a 0.36 nm spacing \cite{Kiang96d}
by analyzing the HRTEM image of individual multi-walled nanotubes.
These reports suggest a spread of inter-shell
distances in carbon nanotubes, with a possible dependence on the tube
size.  In the present work, we study the HRTEM 
nanotube images for several different multi-walled carbon nanotubes
in an effort to understand the 
previously-reported variation in the inter-shell spacing.
To obtain quantitative results, we carried out a digital image 
analysis of HRTEM images, which allows us to relate the 
inter-shell spacings to the nanotube diameter and the number of shells
in a given nanotube.

Carbon nanotube samples were prepared by the usual arc-discharge
method \cite{Iijima91a,Ebbesen92a}.  The core of the deposit was crushed
and dispersed in ethanol.  A drop of this
solution was transferred to a holey carbon microscope grid.
HRTEM images were obtained with a TOPCON 002B microscope at 200KV 
or a JEOL 4000 EX at 400KV accelerating voltage.  
The images were scanned with a CCD camera and stored 
in a 1024 $\times$ 1024 pixel array of 256 gray-scale levels. 

\begin{figure}[htb]
\centering
\leavevmode
\psfig{file=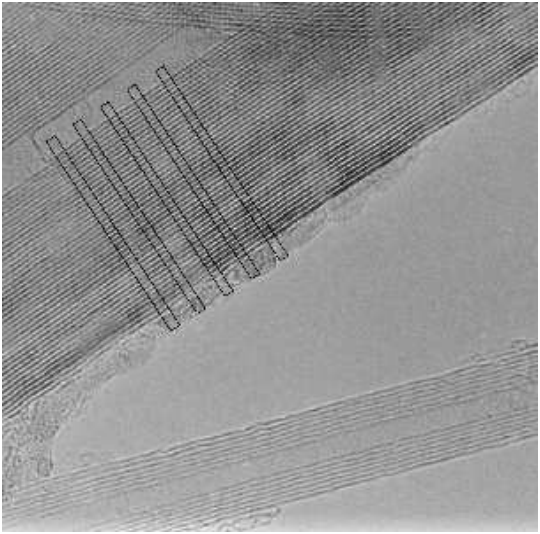,width=2.8in}
\vspace{0.1in}
\caption[TEM]{
High resolution transmission electron microscopy images of
multi-walled carbon nanotubes.  The inter-shell spacing
${\hat{d}_{002}}$ was measured
in the real space images, as indicated by the boxes.}
\label{fig:spcf1}
\end{figure}

\begin{figure}[htb]
\centering\leavevmode
\psfig{file=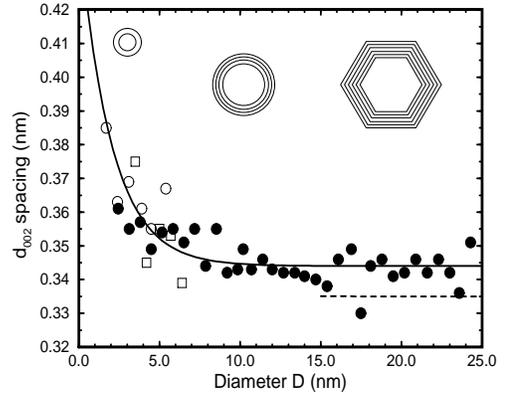,width=2.8in}
\caption[Diameter versus d spacing]
{The spacing ${\hat{d}_{002}}$ decreases as the tube diameter
increases, and approaches 0.344 nm at roughly $D$ = 10
nm.  The data were measured from three different nanotubes indicated 
by different symbols.  Hollow circles: from a 7-shell tube with 
innermost diameter $D_{min}$ = 1.7 nm (shown in Fig.~\ref{fig:spcf1}); 
Solid circles: from a 41-shell tube with $D_{min}$ = 2.6 nm 
(shown in Fig.~\ref{fig:spcf1});
Hollow squares: from a 6-shell tube with $D_{min}$ = 3.5 nm 
(from Ref.~\cite{Kiang96d}, not shown).
The curve ${\hat{d}_{002}} = 0.344 +0.1e^{-D/2}$ 
is the least square fit to the experimental data (see text).
For small tube diameters, the ${\hat{d}_{002}}$ spacing increases exponentially
as the tube diameter decreases.  For intermediate tube diameters,
the ${\hat{d}_{002}}$ spacing is a slowly varying function of tube diameter.
For large tube diameters, graphitization may occur that results
in a polygonal cross-section.  The dashed line indicates
the expected decrease in ${\hat{d}_{002}}$ owing to the 
local graphitic stacking.}
\label{fig:spcf2}
\end{figure}

We carried out high resolution image analysis of nanotubes in real space,
which allows us to measure individual inter-shell
spacings as a function of tube diameter.
Each data point is obtained as an average over 5 measurements to reduce the
error to 3\%, as shown in Fig.~\ref{fig:spcf1}.
%
%
Our data for all tube diameters that were studied
show that the inter-shell spacing (${\hat{d}_{002}}$) ranges 
from 0.34 to 0.39 nm, and that ${\hat{d}_{002}}$ increases as the 
tube diameter decreases.
The empirical equation for the best fit to the data is
\begin{equation}
{\hat{d}_{002}} = 0.344+0.1e^{-D/2}~~~~{\rm for}~D~\ge 0, \label{exp}
\end{equation}
where $D$ is the inner tube diameter, and all the constants are in nm.
Equation~(\ref{exp}) was obtained by a least square fit of the function
${\hat{d}_{002}} = A+B*e^{-C*D}$ to our experimental data (where $A$, $B$,
and $C$ are adjustable parameters), as depicted by the solid curve
in Fig.~\ref{fig:spcf2}.
%
%
The inter-shell spacing decreases exponentially 
and approaches 0.344 nm as the tube diameter increases. 

For small tube diameters ($D <$ 10 nm), 
an increase in the ${\hat{d}_{002}}$ spacing with decreasing tube diameter 
is physically reasonable, since 
the repulsive forces of the graphene basal planes between adjacent tubes 
are larger for smaller diameter tubes, owing to 
their larger curvature, which perturbs
the geometric and electronic structures relative to a planar graphene sheet.
It has been pointed out that the HRTEM projected images of small
diameter carbon naotubes is asymmetric \cite{Cowley96a}, but this effect
on the measured inter-shell spacing is small and is
within our experimental error.
From Eq.~(\ref{exp}) we estimated that the $\hat{d}_{002}$ spacing is 
0.41 nm for a 0.7 nm diameter tube, the smallest 
diameter carbon nanotube observed experimentally \cite{Kiang96c}.
Furthermore, for finite sized tubes, the allowed diameters are discrete, 
which poses constraints on the inter-shell spacings.  
We calculated the distribution of allowed tube diameters with different
helicities according to Eq.~(\ref{D})
\begin{equation}
D ({\rm nm}) = {\sqrt{3} \times 0.142 \over \pi} \sqrt{m^2 + mn + n^2} 
\label{D}
\end{equation}
where $m$ and $n$ are integers.
A discontinuity of 0.01 nm occurs for very small diameter nanotubes 
($D <$~3~nm).  This geometric effect may explain the deviation of
${\hat{d}_{002}}$ values from the predicted curve shown 
in Fig.~\ref{fig:spcf2}.

For large tube diameters ($D >$ 10 nm), the variation in 
${\hat{d}_{002}}$ spacing can be explained by assuming a
uniform charge density
\begin{eqnarray}
{\hat{d}_{002}} & = & \sqrt{(R(R+d_{002}))} 
- \sqrt{(R(R-d_{002}))} \nonumber \\
& \sim & d_{002} + \frac{d_{002}^3}{8R^2} + 
O\left(\frac{1}{R^3}\right)~~as~~ R \rightarrow \infty \label{r2}
\end{eqnarray}
where $R=D/2$ is the tube radius and $d_{002}$ is 0.344 nm, the asymptotic
limit of the ${\hat{d}_{002}}$ spacing.

The asymptotic limit
deduced by Eq.~(\ref{exp}) is 0.344 nm, the same spacing as that
in a turbostratic graphite.  This is expected since defect-free
multi-walled nanotubes have circular cross-sections. 
In any single multi-walled nanotube, the diameter of
each constituent graphene shell is different, which prevents perfectly
correlated graphitic stacking.
Some of the fluctuations in the spacings may be associated
with the jumps in spacing at points where the helix 
angle changes \cite{Cowley94b}.
A $\hat{d}_{002}$ value smaller than that of a turbostratic graphite, 
however, may occur when the energy gained by the local graphitic 
stacking is less than the energy introduced by the defects 
associated with the stacking.
Inter-shell spacings smaller than 0.344 nm,
resulting from a polygonal cross-section, have frequently been observed 
for large diameter nanotubes \cite{Cowley94a,Ebbesen94a} and 
for vapor-grown carbon fibers \cite{Dresselhaus89a}.

\begin{figure}[htb]
\centering
\leavevmode
\psfig{file=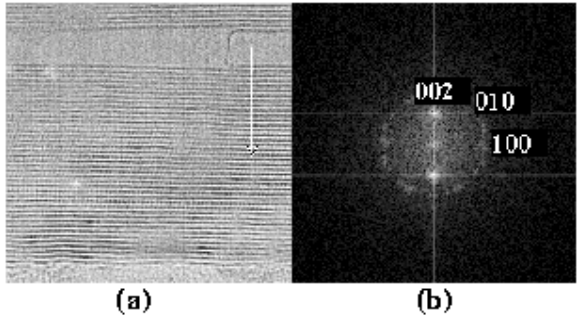,width=3in}
\vspace{0.01in}
\caption[Real space image and FFT.]
{(a) Digitized high resolution transmission electron micrograph 
of the core region of the top carbon nanotube shown in 
Fig.~\ref{fig:spcf1}, where the \{100\} lattice 
fringes are clearly resolved.  
The average diameter is illustrated in this micrograph.
(b) Fast Fourier transform (FFT) of the nanotube.  
The (002), (100), and (010) reflections are resolved.  
The position of the spots is measured at the location of the
highest intensity.
The $d_{100}$ spacing (0.213 nm) is used as a calibration standard.}
\label{fig:spcf3}
\end{figure}

We also measured the lattice constants in reciprocal space, 
which complement the results obtained from real space images.
With a known minimum diameter of a nanotube (measured directly from the
HRTEM images), we calculated the diffracted spot positions
from a crystal with a slowly-varying $c$ unit cell length defined by
Eq.~(\ref{exp}).  
The calculated results were then compared to the experimental data.
Figure~\ref{fig:spcf3}a presents data for a nanotube with 
the 0.213 nm $d_{100}$ lattice 
fringes of the graphite basal plane clearly resolved, and
the corresponding FFT is shown in Fig.~\ref{fig:spcf3}b. 
%
%
The inter-shell spacings 
(${\hat{d}_{002}}$) were deduced from the power spectra, 
using the location of the highest intensity as the 
spot position, as shown in Fig.~\ref{fig:spcf3}b

We used the 0.213 nm spacing between \{100\} fringes 
as a calibration standard, based on 
the X-ray diffraction study \cite{SaitoY93d} showing that
the C-C bond lengths in the nanotubes are
the same for both graphite and carbon nanotubes (0.142 nm)
\cite{SaitoY93d}.  The tube diameters, defined as 
the average diameter ($D_a$) of the tubes used for the FFT, 
were obtained from the real space images.

\begin{figure}[htb]
\centering
\leavevmode
\psfig{file=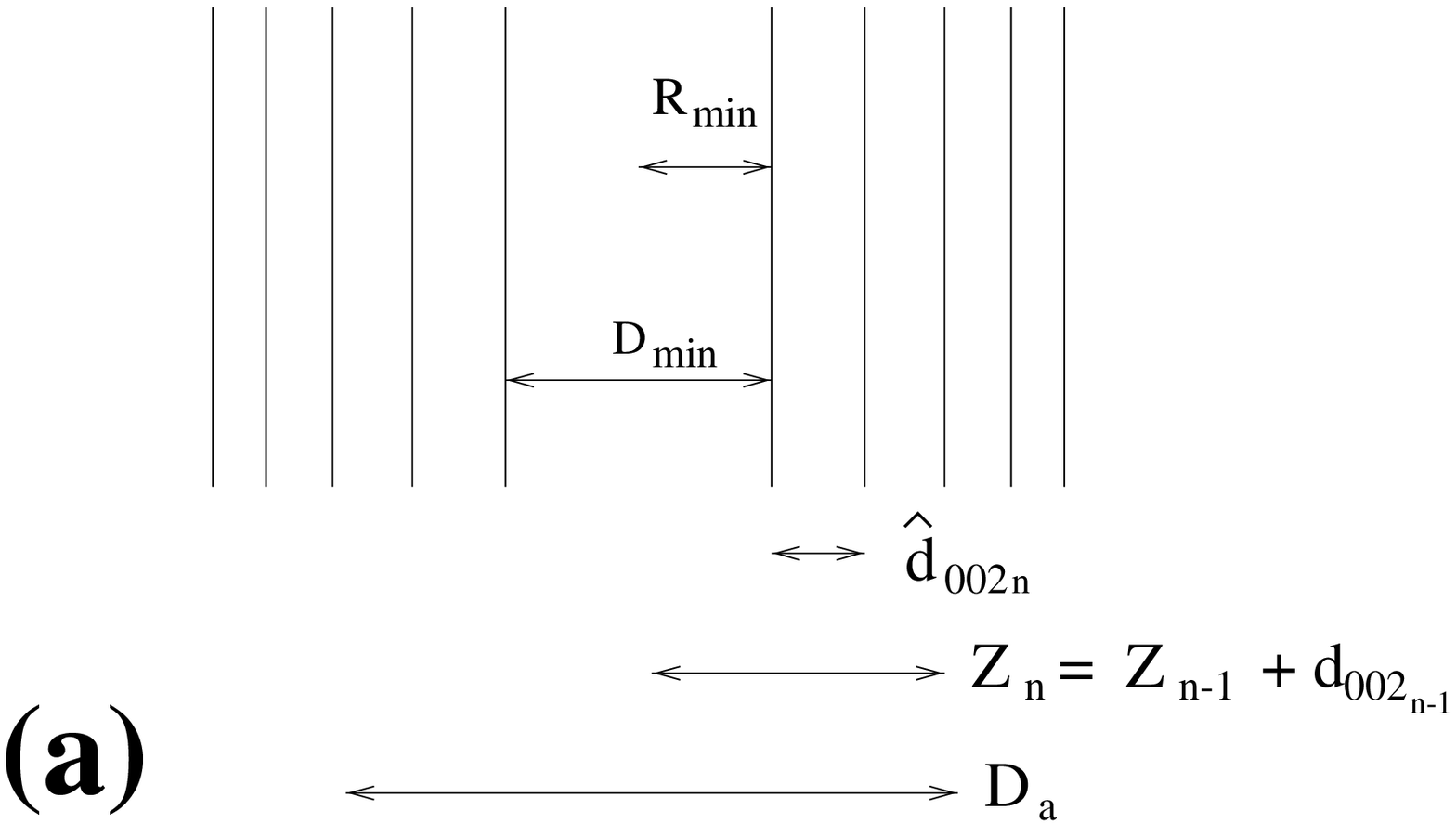,width=2.4in}
\psfig{file=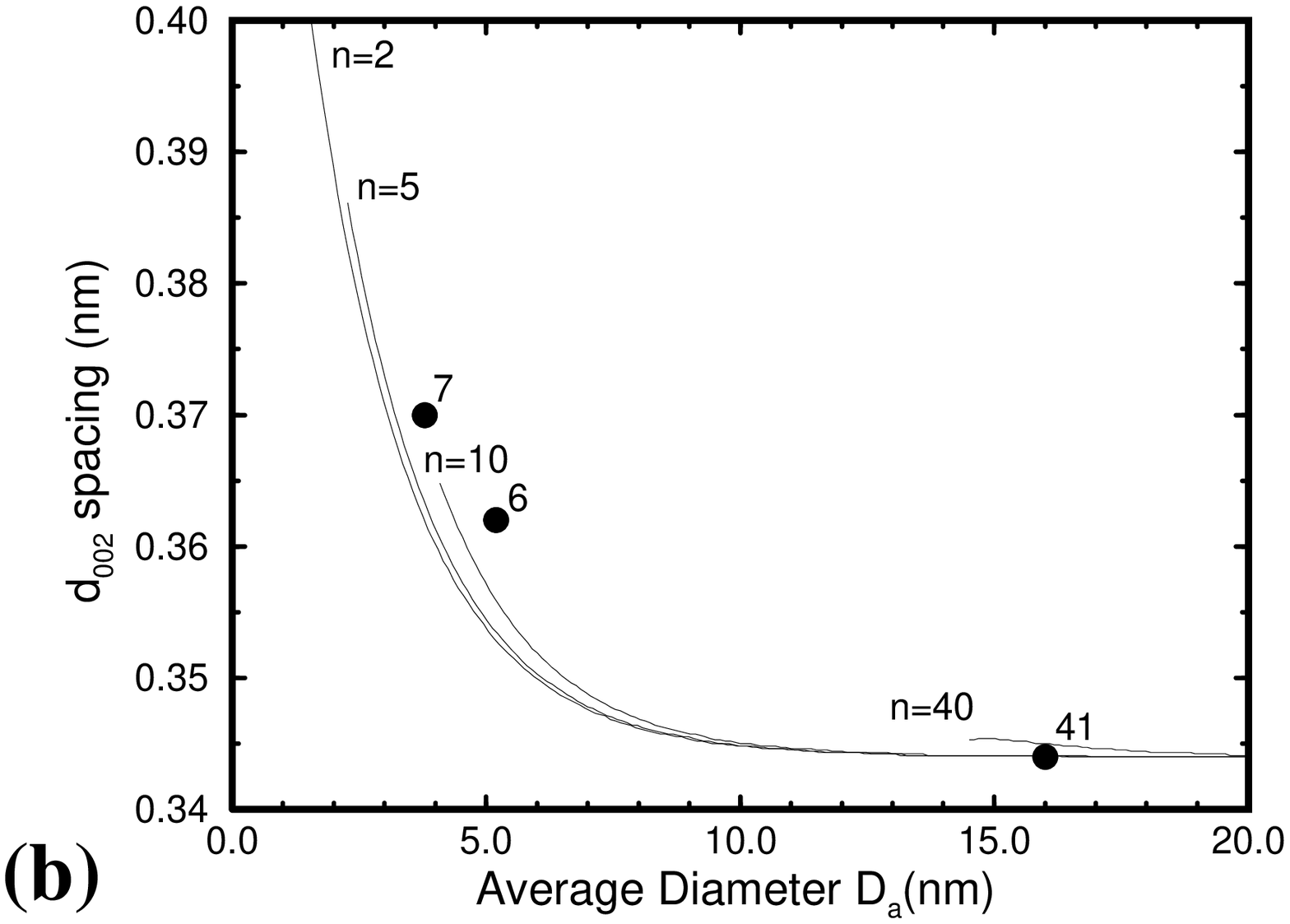,width=2.8in}
\caption[Model]{
(a) Model for a nanotube crystal with a varying inter-shell spacing.  
The ${\hat{d}_{002}}$ as an exponential function of tube diameter 
is defined in Eq.~(\ref{exp}).
(b) The inter-shell spacing ${\hat{d}_{002}}$ as a function of the 
average tube diameter ($D_a$), where
n is the number of shells in a nanotube.
The curves are calculated results for n= 2, 5, 10, and 40,
using the above model and the exponential function from Eq.~(\ref{exp}).
The three data points (with 2\% error) shown by the large full circles
were obtained from the power spectra of the 
two tubes surveyed in Fig.~\ref{fig:spcf1} 
and the tube in Ref.~\cite{Kiang96d}.}
\label{fig:spcf4}
\end{figure}

In reciprocal space, the scattering amplitude $F(k)$ is 
\begin{equation}
F(k) =  \sum_n f(k) e^{-i{\bf k \cdot r_n}}, \label{amp}
\end{equation}
where $n$ is the $nth$ unit cell in the crystal, ${\bf k}$ is the reciprocal
lattice vector, ${\bf r}$ is the atomic position in real space,
and $f(k)$ is proportional to the atomic form factor for electron 
scattering of carbon and is a slow varying function of $k$ 
for electron scattering.
Since we were only interested in diffraction along 
the (002) direction, we constructed a crystal of parallel graphene sheets
with varying inter-shell distances, as depicted in Fig.~\ref{fig:spcf4}a.
%
%
Defining $z$ to be perpendicular to the graphene planes, we can 
express $z$ as a function of the cell $n$ (for $n \geq 2$)
\begin{equation}
z_n = z_{n-1} + {\hat{d}_{002_{n-1}}}, \label{z}
\end{equation}
where $z_n$ is the atomic position 
of the $nth$ unit cell along the $z$-axis relative to the origin, 
${\hat{d}_{002}}$ is evaluated with Eq.~(\ref{exp}),
and $z_1=R_{min}$ is the minimum radius of the nanotubes.
The scattering amplitude $F$ 
for the crystal shown in Fig.~\ref{fig:spcf4}a is 
\begin{equation}
F = f \sum_{n=1}^{N} e^{-i k_z z_n}, 
\end{equation}
where $N$ is the total number of unit cells along the $z$ axis. 
Thus the diffracted intensity $I$ is 
\begin{equation}
I = \left\vert F \right\vert ^2 = 
f^2 \left[\left(\sum_{n=1}^{N} cos(k_z z_n)\right)^2 
+ \left(\sum_{n=1}^{N} sin(k_z z_n)\right)^2\right].    \label{int}
\end{equation}

For a crystal with a constant ${\hat{d}_{002}}$, Eq.~(\ref{int}) can be 
evaluated analytically, and the peak position 
(defined as the intensity maximum of the peak) is at
$2\pi/{\hat{d}_{002}}$.
For a crystal with a slowly varying 
${\hat{d}_{002_n}} = {d}_{002}+\delta d(n)$ for all tubes, 
the peak is shifted relative to the $k_z$ value for ${d}_{002}$.   
Using this model, we evaluated numerically the ${\hat{d}_{002}}$ spacing as 
a function of the average tube diameter ($D_a$),
starting with the smallest possible $D_a$ value which can occur
for a given $n$, where $n$ is the number of shells in a nanotube. 
By knowing the number of shells and the average diameter of a nanotube,
we can predict the (002) spot position in Fourier space.
The resulting curves for $n$ = 2, 5, 10, and 40 are shown
in Fig.~\ref{fig:spcf4}b.   
The data from the Fourier analysis of 
the three tubes surveyed in Fig.~\ref{fig:spcf2} (where $n$ = 6, 7, and 41)
agree with our model, supporting our conclusion that the 
inter-shell spacing is a function of tube diameter.

We also found that the nanotubes with smaller innermost
tube diameters often have fewer shells.
This may be because that the kinetics of the growth reaction is 
correlated with the product stability.
According to the multi-walled nanotube growth mechanism 
\cite{Endo92a,Iijima92b}, 
nanotubes grow by first forming a nucleating shell, followed
by epitaxial growth of additional shells.  
Larger inter-shell spacings resulting from a higher strain 
energy of the nanotubes decreases the stability of the nanotube.
Because of the larger inter-shell
spacings and lower stability of small diameter tubules, 
the epitaxial growth of additional shells 
onto the nucleating core is slower for smaller diameter tubes in
comparison with larger diameter tubes.
Therefore, the radial growth rate of nanotubes is 
also expected to be related to the nucleating tube diameter.

The small number of shells $n$ may also contribute to the size effect
that causes increased inter-shell spacing.  
This effect, however,
is expected to be smaller compared to the diameter effect.
According to the shell-by-shell growth model, 
a variation of the inter-shell spacing as a function of $n$ requires
a change in the inter-shell spacing as the number of shells increases, 
which would result in a decrease in C-C bond length.  
For example, a spacing that changes from 0.37 to 0.35 nm
will result in a 5\% decrease in bond length, which implies
reducing the C-C bond length from 0.142 nm to 0.134 nm,
a nearly impossible process.
Our experimental results can be explained solely by the diameter size effect, 
which suggests that the effect due to a small number of carbon 
shells is within the experimental uncertainties.  
Incorporation of the $n$ effect may, 
however, provide further refinement to our model.

In summary, we have studied the nanotube
structures by high resolution transmission electron microscopy
and digital image analysis.
We found that the inter-shell spacing
of nanotubes decreases with increasing tube diameter,
approaching $0.344$~nm asymptotically at a tube diameter of roughly
10 nm.  The size effect is more profound
in the small diameter ($D <$~10~nm) region, where each additional 
shell results in a measurably different ${\hat{d}_{002}}$ spacing.  
The physical and chemical properties may also vary 
owing to the change in inter-shell spacing of the nanotubes. 
For example, nanotubes of larger inter-shell spacing should be less 
stable and, therefore more reactive.
The variation in interaction forces of small diameter nanotubes 
should modify single-walled carbon nanotube 
surface properties, which may result in 
useful characteristics for storage media such as hydrogen 
fuel cells and batteries.

We thank Dr.\ S.\ Iijima for helpful discussions.
CHK and PMA acknowledges the financial support 
from the UC Energy Institute and the Alexander von Humboldt
Foundation, respectively.
Part of the work by ME is supported by the Ministry of
Education, Science and Culture, Japan. 
The work at MIT is supported by NSF grant DMR-95-10093.

\vspace*{-0.3in}
\bibliography{spcrep}
\end{document}